\documentclass[twocolumn,prb,showpacs,preprintnumbers,amsmath,amssymb]{revtex4}

\usepackage[pdftex]{graphicx}
\usepackage{dcolumn}
\usepackage{bm}
\usepackage{amssymb}
\usepackage{wasysym}
\usepackage{sidecap}
\usepackage{bibtopic}

\def\kperp{{\bf k}$_\perp$}

\def\kperp{{\bf k}$_\perp$}

\def\Kbar{$\overline{\rm K}$}
\def\Gbar{$\overline{\Gamma}$}
\def\Mbar{$\overline{\rm M}$}
\def\GbarMbar{$\overline{\Gamma}$-$\overline{\rm M}$}
\def\GbarKbar{$\overline{\Gamma}$-$\overline{\rm K}$}

\def\BiTe{Sb$_2$Te$_3$}
\def\BiTe{Bi$_2$Te$_3$}
\def\BiSe{Bi$_2$Se$_3$}

\begin{document}

\title{Anisotropic effect of warping on the lifetime broadening of topological\\ surface states in angle-resolved photoemission from Bi$_2$Te$_3$}

\author{J. S\'anchez-Barriga$^1$, M. R. Scholz$^{1,2}$, E. Golias$^1$, E. Rienks$^1$, D. Marchenko$^{1,3}$,  A. Varykhalov$^1$, L. V. Yashina$^4$ and O. Rader$^1$}

\affiliation{$^1$Helmholtz-Zentrum Berlin f\"ur Materialien und Energie, Elektronenspeicherring BESSY II, Albert-Einstein Str. 15, 12489 Berlin, Germany}
\affiliation{$^2$Physikalisches Institut und R\"ontgen Center for Complex Materials Systems, Universit\"at W\"urzburg, 97074 W\"urzburg, Germany}
\affiliation{$^3$Physikalische und Theoretische Chemie, Freie Universit\"at Berlin, Takustr. 3, 14195 Berlin, Germany}
\affiliation{$^4$Department of Chemistry, Moscow State University, Leninskie Gory, 1/3, 119992 Moscow, Russia}

\date{\today}

\begin{abstract}
We analyze the strong hexagonal warping of the Dirac cone of \BiTe\ by angle-resolved photoemission. Along \Gbar\Mbar, the dispersion deviates from a linear behavior meaning that the Dirac cone is warped outwards and not inwards. We show that this introduces an anisotropy in the lifetime broadening of the topological surface state which is larger along \Gbar\Kbar. The result is not consistent with nesting. Based on the theoretically predicted behavior of the ground-state spin texture of a strongly warped Dirac cone, we propose spin-dependent scattering processes as explanation for the anisotropic scattering rates. These results could help paving the way for optimizing future spintronic devices using topological insulators and controlling surface-scattering processes via external gate voltages.
\end{abstract}


\maketitle

Topological insulators (TIs) are characterized by an insulating bulk energy gap and gapless spin-polarized Dirac-cone surface states with electron spins locked perpendicular to their linear momenta.\cite{Fu-PRL-2007,Hasan-RMP-2010} This peculiar spin texture is believed to play a central role in inducing exotic quantum phenomena \cite{Hasan-RMP-2010, KanePRL2005}, novel magnetic-spin physics,\cite{Essin, Zhang} as well as in the development of future spin-based low-power transistors among a variety of applications. Such perpendicular locking can be realized if spins of electrons occupying the Dirac cone move around in ideally circular constant-energy contours between the Dirac point and the Fermi level.\cite{Hsieh-Nature-2009} Owing to the spin-momentum locking, electrons on the surfaces of TIs are protected from backscattering,\cite{Roushan} an effect that might be of crucial importance in the generation of spin currents with reduced dissipation,\cite{Zhang-Physics-2008} coherent spin rotation,\cite{Nowack} spin-orbit qubits,\cite{Kouwenhoven-Nature-2010} as well as in other applications such as manipulation of photon-polarization driven spin currents in real devices.\cite{Gedik_NT} The spin texture of the Dirac cone can be affected by the presence of hexagonal warping,\cite{Fu:2009p10385}
 i.e., the Dirac cone is deformed in such a way that from the Dirac point to the energy of the Fermi level the constant-energy contours develop from an ideally circular shape to a hexagon, and subsequently to a snowflake-like Fermi surface. This strong distortion of the Dirac cone might open up new possibilities for observing other interesting phenomena, such as the breaking of time-reversal symmetry and the surface quantum Hall effect under applied magnetic fields parallel to the surface,\cite{Fu:2009p10385} enhanced surface scattering or the formation of surface spin-density waves.\cite{Fu:2009p10385,Hasan:2009p10386}

In the context of exploring these possibilities, the observation and importance of hexagonal warping in the band dispersions of the topological surface states (TSSs),\cite{Chen:2009p10295,Kuroda:2010p11138,Hirahara:2010p11227} as well as its theoretical description,\,\cite{Fu:2009p10385,Frantzeskakis:2011p13370,Basak:2011p13107} has stimulated a manifold of theoretical and experimental investigations on TIs. The strong influence of warping on the spin texture of TSSs giving rise to a finite out-of-plane spin orientation was theoretically predicted,\,\cite{Fu:2009p10385} and by means of spin-resolved angle-resolved photoemission (SR-ARPES) experimentally observed.\,\cite{Souma:2011p13166} That in the presence of warping the electron spins are not anymore locked perpendicularly to their linear momentum was also proposed,\,\cite{Basak:2011p13107} and the lifting of such perpendicular spin-momentum locking systematically investigated by means of SR-ARPES experiments on TSSs of the prototypical TIs \BiTe\ and \BiSe.  \cite{Hsieh-Nature-2009,Souma:2011p13166, Pan11,Jozwiak11,JozwiakNP13,Sanchez-Barriga-PRX-2014, Pan-PRB-2013} These investigations revealed that in \BiTe, where warping effects are much stronger than in \BiSe,\cite{Souma:2011p13166} the distorted Dirac cone is characterized by a left-handed three-dimensional spin texture which features an out-of-plane spin component oscillating around the Fermi surface, and an in-plane spin component tangentially following the snowflake contour which results from strong warping.\cite{Souma:2011p13166}  

In a similar way, the change that warping induces in the ARPES signal obtained from circular dichroism in the angular distribution was attributed to the lifting of the perpendicular spin-momentum locking,\,\cite{Wang:2011p13090} or alternatively to the strong influence that warping induces on the orbital angular momentum of TSSs.\,\cite{Jung:2011p13108} The influence of warping on the spin orientation was additionally predicted to alter the possible channels for quasiparticle scattering, and this effect has been the central topic of various theoretical investigations on TSSs taking into account spin-orbit scattering\,\cite{Lee:2009p13114} or scattering at magnetic point defects.\,\cite{Zhou:2009p13118} The effect of quasiparticle interference in TIs was additionally studied by means of scanning tunneling microscopy (STM) and spectroscopy (STS) and from this type of experiments, an enhanced magnitude of the surface patterns resulting from electronic interference was observed and ascribed to hexagonal warping. \,\cite{Zhang:2009p10845,Okada:2011p13133,Kim:2011p13203,Wang:2011p13355} Moreover, the possibility of a spin-density wave and Friedel-like oscillations of the local density of states was predicted\, \cite{Fu:2009p10385} and the latter experimentally observed by means of STM experiments.\,\cite{Alpichshev:2010p13116} Equally important, a current-induced spin polarization was theoretically predicted and it was shown that such a current may enhance the spin component perpendicular to the surface when significant hexagonal warping is present.\,\cite{Misawa:2011p13369,Wang:2011p13089} 

In view of these observations and predictions, particularly taking into account that warping can influence the spin orientation of the surface-state electrons\,\cite{Fu:2009p10385,Basak:2011p13107,Souma:2011p13166} and that the spin orientation might alter the scattering properties of the surface electrons,\,\cite{Lee:2009p13114,Zhou:2009p13118} it is critically important to especially investigate the momentum-dependence of the lifetime broadening of TSSs in the presence of strong hexagonal warping. Namely, whether a large Fermi-surface distortion is a weak perturbation on the lifetime broadening of TSSs or there are preferential momentum-space directions where this distortion results into a substantial modification of the scattering rates of the surface-state electrons. ARPES represents an ideal technique for such a purpose, as it is a well-established method to determine the lifetime broadening of electronic states,\cite{Smith-PRB-93, Damascelli-PS-2004} not only in strongly correlated systems such as transition-metal oxides,\cite{Wadati-LNP-2007} 4$f$ rare-earths,\cite{Sekiyama-LNP-2007} high-$T_{c}$ superconductors \cite{Fink-LNP-2007, Valla:1999p13362} or 3$d$ ferromagnets, \cite{Sanchez-Barriga-PRL-2009,Sanchez-Barriga-PRB-2010,Sanchez-Barriga-PRB-2012} but also in TIs.\cite{Valla:2012p13194, Pan-PRL-2012, Chen-SciRep-2013} Moreover, ARPES has recently become the most powerful tool in systematically revealing the linear dispersion of TSSs in energy and momentum space.

In the present work, we utilize the TSS of the prototype TI \BiTe\ to analyze the strong hexagonal warping of the Dirac cone by angle-resolved photoemission. We investigate the impact of warping on the scattering properties of the surface-state electrons in this system. We demonstrate that the large distortion introduced by warping leads to an anisotropy in the lifetime broadening of the TSS which is larger along the \Gbar\Kbar\ direction of the surface Brillouin zone (SBZ). We further identify the underlying mechanism that gives rise to anisotropic scattering rates of the surface-state electrons. This result could help understanding how to control surface-scattering processes that are relevant  for spin injection in future spintronic devices based on TIs.

We perform ARPES measurements at a temperature $T=$30 K using $p+s$ linearly-polarized undulator radiation in ultrahigh vacuum better than $1\cdot10^{-10}$ mbar with a Scienta R8000 electron analyzer at the UE112-PGM2a beamline of BESSY II. \BiTe\ single crystals are grown by the Bridgman method and cleaved {\it in situ}, with the sample temperature kept at 30 K. We use 21 eV and 55 eV photons in combination with the sample geometry shown in Fig. 1(a), where emitted photoelectrons are collected along the electron wave vector {\bf k}$_{e}$ at an angle $\phi=45^{\circ}$ with respect to the incident photon beam, which lies along the \Gbar\Kbar\ or \Gbar\Mbar\ directions of the SBZ within the yz plane of the laboratory reference frame. ARPES spectra are taken by rotating the sample about the $x$ axis or $z$ axis, while keeping the rotation about the $y$ axis fixed. We use an analyzer entrance slit size of 0.2 mm $\times$ 25 mm in width and length, respectively. Note that in the laboratory reference frame, ${\bf k}_{\parallel,x}$ and ${\bf k}_{\parallel,y}$ wave vectors are fixed parallel and perpendicular to the analyzer entrance slit, respectively. The long axis of the entrance slit is parallel to the electron emission plane [red (light) dashed lines in Fig. 1(a)]. The angular resolution of the ARPES experiment parallel and perpendicular to the slit is about 0.1$^{\circ}$, and the energy resolution $\sim$30 meV. The high quality of the achieved (111) surfaces is verified by the sharp features in the low-energy electron diffraction (LEED) patterns [see Fig. 1(b)] and in angle-resolved photoemission of the valence band [see Fig. 2].
\begin{figure}
\centering
\includegraphics [width=0.49\textwidth]{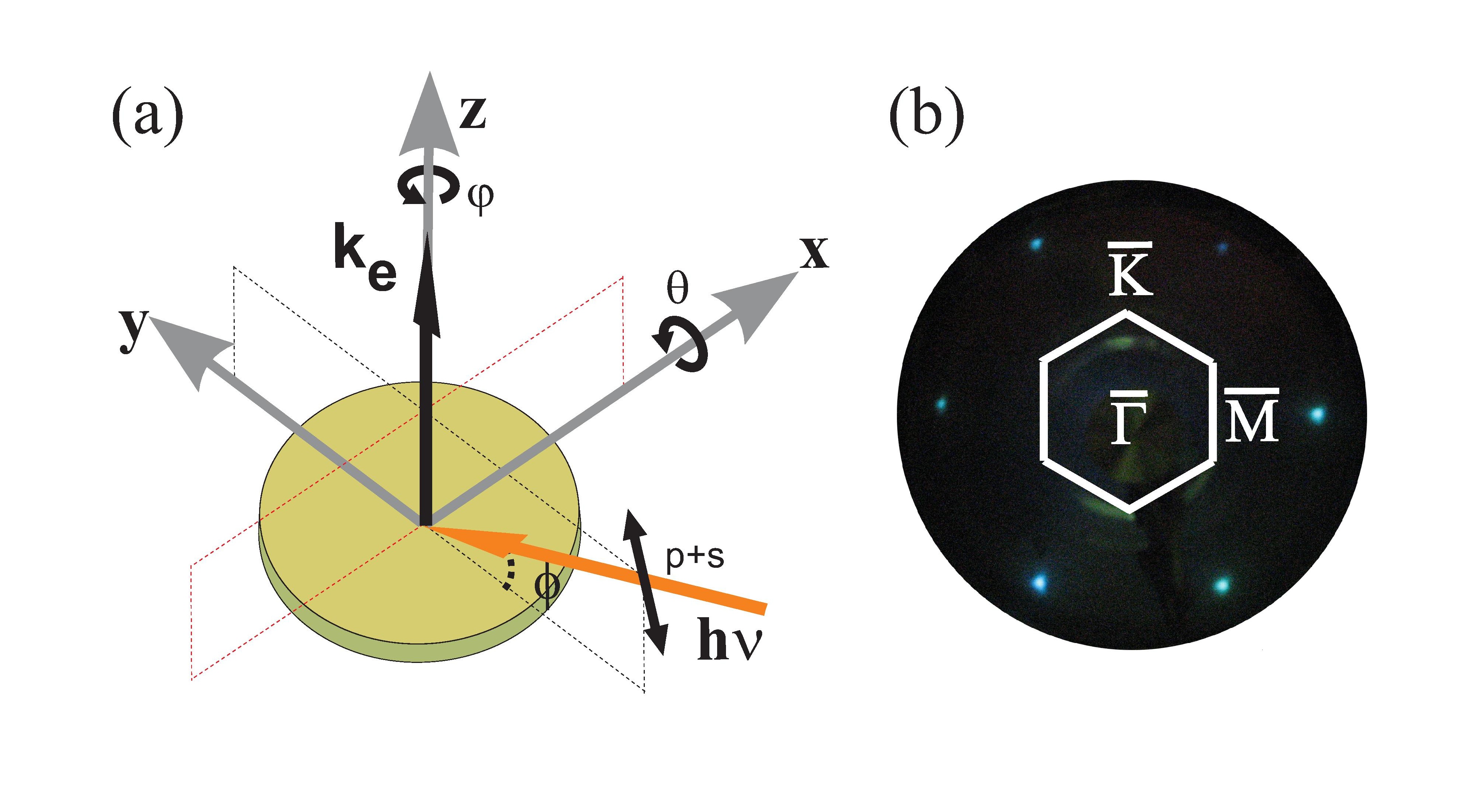}
\caption{(Color online). (a) Geometry of the ARPES experiment in the laboratory reference frame. The light incidence and electron emission planes are indicated by black (dark) and red (light) dashed lines, respectively. (b) LEED pattern of the \BiTe\ (111) surface acquired with an electron beam of 25 eV.}
\label{Characterization1}
\end{figure}
 
\begin{figure*}
\centering
\includegraphics [width=0.9\textwidth]{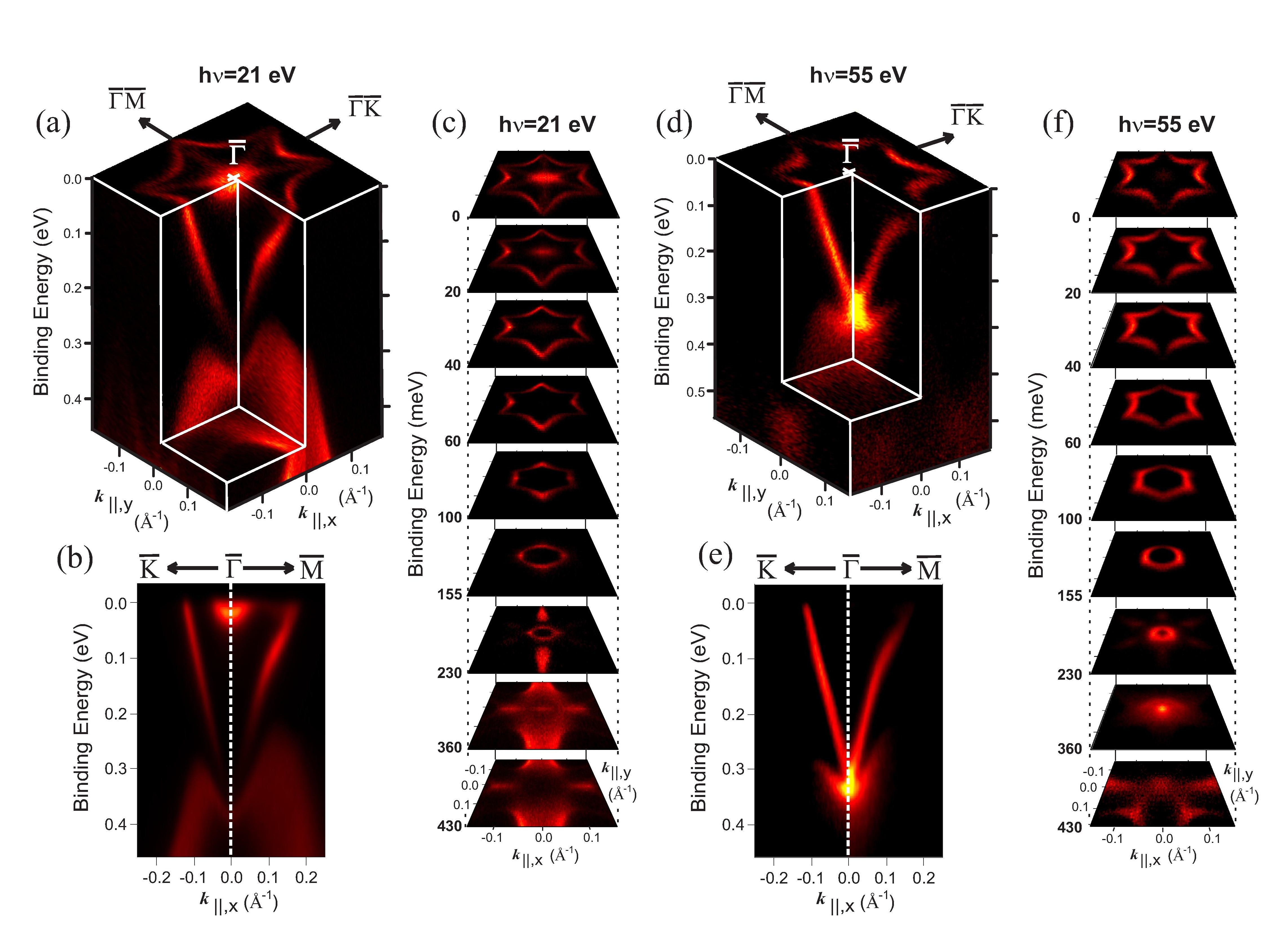}
\caption{(Color online). (a) High-resolution ARPES dispersions of the topological surface state, bulk conduction band and bulk valence band states of \BiTe\ sampled using 21 eV photons around the \Gbar\ point of the surface Brillouin zone along both in-plane ${\bf k}_{\parallel,x}$ and ${\bf k}_{\parallel,y}$ momenta. (b) Comparison between the ARPES dispersions obtained along \Gbar\Mbar\ and \Gbar\Kbar. Both high-symmetry directions were oriented parallel to ${\bf k}_{\parallel,x}$. (c) Constant-energy surfaces extracted at different binding energies from the full photoemission mapping in (a). (d),(e),(f) Analogous results to those obtained in (a),(b),(c), respectively, but using a photon energy of h$\nu$=55 eV.}
\label{Characterization2}
\end{figure*}

Figure 2(a) displays high-resolution ARPES dispersions of the TSS, bulk conduction band (BCB) and bulk valence band (BVB) states of \BiTe\ sampled with 21 eV photons around the \Gbar\ point of the SBZ along both in-plane ${\bf k}_{\parallel,x}$ and ${\bf k}_{\parallel,y}$ momenta. The differences in the electron dispersions for the two high-symmetry directions \Gbar\Mbar\ and \Gbar\Kbar\ sampled along ${\bf k}_{\parallel,x}$ are emphasized in Fig. 2(b). Figure 2(c) displays the corresponding constant-energy surfaces extracted at different binding energies from the full photoemission mapping in Fig. 2(a). Figures 2(d), 2(e), and 2(f) display analogous results to those obtained in Figs. 2(a), 2(b) and 2(c), respectively, but using a photon energy of h$\nu$=55 eV. Quantization effects due to band bending\,\cite{Bianchi-NatComm-2011, King-PRL-2011, Bianchi-PRL-2011} are absent, indicating that our ultrahigh vacuum experimental chamber contains an extremely small amount of residual gas which is not sufficient to generate a pronounced band bending effect giving rise to quantum-well states. A gapless Dirac cone representing the TSS with a Dirac point located at a binding energy (BE) of $\sim$360 meV is clearly observed. The BE position of the BCB crossing the Fermi level, which at 21 eV reaches its minimum at a BE of $\sim$35 meV [see Fig. 2(c)], indicates that the crystals are intrinsically $n$-doped. The intensity of the Dirac point strongly changes with the photon energy due to photoemission matrix elements,\cite{Kordyuk-PRB-2011} which also lead to changes in the relative intensities from the TSS and bulk bands at different photon energies [compare, e.g., Figs. 2(a) and 2(d)]. Moreover, because in ARPES the photon energy selects the component of the electron wave vector perpendicular to the surface \kperp, the contribution from bulk states to the ARPES intensity at a particular photon energy is also affected by their characteristic \kperp dispersion. As a result of these effects, the BCB crossing the Fermi level, which clearly contributes at 21 eV, is strongly suppressed and virtually disappears at 55 eV photon energy [Figs. 2(d)-2(f)]. In contrast, the energy-momentum dispersion of the TSS, despite the additional changes in its intensity distribution, remains clearly the same when varying the photon energy from 21 eV to 55 eV [compare, e.g., Figs. 2(b) and 2(e)]. 

The TSS is strongly distorted by hexagonal warping. As a result, the constant-energy contours shown in Figs. 2(c) and 2(f) progressively evolve from an ideal circular shape between the Dirac point and a BE of $\sim$190 meV, to a hexagon up to a BE of $\sim$110 meV, to progressively more pronounced snowflake-like constant-energy contours in a BE range of about 100 meV below the Fermi level. On the other hand, the lower Dirac cone, which emerges once the TSS bands disperse back to lower BE once they cross at the Dirac point, follows the energy-momentum dispersion of the BVB, which at 21 eV reaches its maximum at a BE of $\sim$230 meV along the \Gbar\Mbar\ direction of the SBZ [see Fig. 2(b)]. As a result, an overlap between the lower Dirac cone and BVB can be observed at this photon energy. This is also the case at 55 eV, where the intensity from the BVB, although weaker, clearly contributes to the measured ARPES intensity, in contrast to the BCB which is strongly suppressed [see Figs. 2(d)-2(f)]. At 21 eV photon energy, where the BCB intensity is very large, a strong overlap between BCB states and the TSS can be clearly distinguished along the \Gbar\Mbar\ direction in a narrow BE range of $\sim$20 meV below the Fermi surface [see, e.g., Figs. 2(a) and 2(c)]. This overlap is due to the fact that the BCB is also affected by the hexagonal distortion which follows the three-fold symmetry of the crystal structure around the [111]-direction. In consequence, the BCB extends towards the TSS snowflake-like Fermi surface and touches it only along the \Gbar\Mbar\ directions of the SBZ. We note that the BVB also exhibits a distorted three-fold symmetric pattern but with somewhat different intensity distribution. In spite of this distortion, only the lower Dirac cone overlaps with the BVB, and this overlap seems to extend over a wide BE range, from around the Dirac point towards higher BE values. 

In order to quantitatively determine the changes in the lifetime broadening of the TSS in the presence of hexagonal warping, in the following we perform a detailed analysis of the TSS spectral width using momentum distribution curves (MDCs) extracted from the ARPES dispersions shown in Figs. 2(b) and 2(e), where both \Gbar\Kbar\ and \Gbar\Mbar\ high-symmetry directions of the SBZ are oriented along ${\bf k}_{\parallel,x}$ wave vectors (i.e., parallel to the analyzer entrance slit). We use the standard method where the MDCs are fitted with Lorentzian peaks convoluted with a Gaussian function representing the momentum resolution.\,\cite{Valla:1999p13362, Valla:1999p13376, Valla:2012p13194} The results of our analysis using this procedure are shown in Fig. 3, and are only valid under the assumption that TSSs are weakly-interacting states. Note that in our evaluation the scattering rates of the surface-state electrons $\Gamma=2\Sigma''=v_{0}\Delta{\bf k}(E)$ are directly related to the half width at half maximum (HWHM) of the Lorentzian
peaks ($\Delta{\bf k}(E)/2$), where $\Sigma''=$Im$\Sigma({\bf k},E)$ is the imaginary part of the complex self-energy $\Sigma({\bf k},E)=$Re$\Sigma({\bf k},E)+i$Im$\Sigma({\bf k},E)$, and $v_{0}$ is the bare group velocity. This  assumption is valid under the approximation in which $v_{0}$ is the renormalized group velocity $v_{g}$, implying that the real part of the self-energy
$\Sigma'$ is small, in agreement with recent observations on TSSs where a weak renormalization due to electron-phonon coupling in the vicinity of the Fermi level and slowly varying energy-dependent values of $\Sigma''$ have been found.\cite{Valla:2012p13194, Chen-SciRep-2013} We point out that due to such a BE dependence of $\Sigma''$,\cite{Valla:2012p13194, Chen-SciRep-2013} the surface-state peaks still show a slight asymmetry with increasing BE, but by fitting MDCs with convoluted Lorentzian functions we have nevertheless obtained accurate results.

\begin{figure*}
\centering
\includegraphics [width=0.95\textwidth]{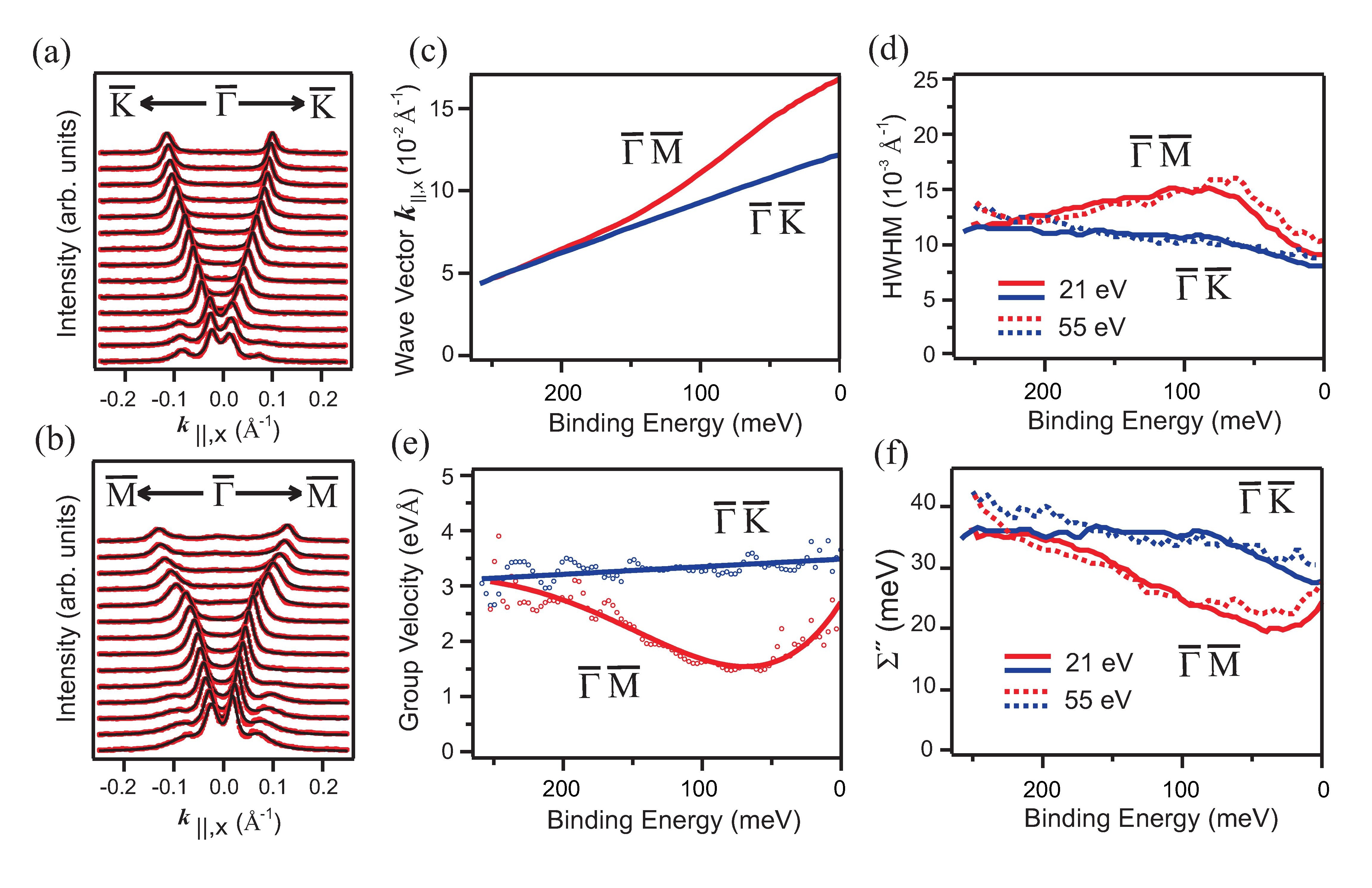}
\caption{(color online). Analysis of MDCs of the surface state of \BiTe. (a),(b) Few-selected fits [black (dark) solid lines] to experimental MDCs [red (light) dotted lines] obtained by varying the BE along (a) \Gbar\Mbar\ and (b) \Gbar\Kbar\ directions, respectively, and using 55 eV photons. (c) The ${\bf k}_{\parallel,x}(E)$ dispersion of the TSS along the \Gbar\Mbar\ and \Gbar\Kbar\ directions  [red (light) and blue (dark) solid lines, respectively] shows an anisotropic behavior. (d) The anisotropy is also reflected in the HWHM of the Lorentzian peaks, which for 21 and 55 eV photons provide similar results (solid and dotted lines, respectively). (e) The group velocity, obtained from a polynomial fit (solid lines) to the derivative of the $E({\bf k})$ dispersion (open circles), is strongly altered over the analyzed BE range along the \Gbar\Mbar\ direction (eV\AA$\equiv\hbar$m/s). (f) Multiplication of the HWHM with the group velocity at each BE gives the imaginary part of the self-energy $\Sigma''$, which reflects scattering rates that are anisotropic in ${\bf k}$-space.}
\label{Anisotropy1}
\end{figure*}

Figures \,\ref{Anisotropy1}\,(a) and \,\ref{Anisotropy1}\,(b) show few-selected fits [black (dark) solid lines] to experimental MDCs [red (light) dotted lines] obtained by varying the BE along \Gbar\Mbar\ and \Gbar\Kbar\ directions, respectively, and using 55 eV photons. Similar results were obtained at 21 eV by fitting the additional contribution from the BCB with an extra Lorentzian peak introduced into the analysis procedure, in similar way as previous reports at 8 eV photon energy for aged-Bi$_2$Se$_3$ surfaces.\cite{Park:2010p10834} Figure \,\ref{Anisotropy1}\,(c) shows the corresponding surface-state ${\bf k}_{\parallel, x}(E)$ dispersions extracted from the fitted peak BE positions along the \Gbar\Mbar\ and \Gbar\Kbar\ directions, which are represented by red (light) and blue (dark) solid lines, respectively. Along \Gbar\Kbar, the dispersion differs only slightly from a linear behavior. We have obtained accurate fits from the Fermi level up to BE values of $\sim$260 meV. In the case of the lower Dirac cone, the above mentioned contribution from the BVB at 21 and 55 eV prevents an accurate fitting. The different behavior of the fitted dispersions along the two high-symmetry directions, as  seen in Fig.\,\ref{Anisotropy1}\,(c), is in agreement with the presence of hexagonal warping. However, our experimental results are not consistent with the expectations of $k\cdot p$ theory, which predicts a strong deviation from the linear dispersion with increased group velocity along the \Gbar\Kbar\ direction.\,\cite{Fu:2009p10385} In reality, we observe a deviation to reduced group velocities along \Gbar\Mbar, meaning that the Dirac cone is warped outwards and not inwards along this direction. Such a disagreement between $k\cdot p$ theory and experiment is due to the theoretical energy-momentum dispersion, despite the good agreement with the experimental Fermi-energy contour. The origin of the discrepancy comes from the effect of bulk conduction bands, which come down in energy along the \Gbar\Mbar\ direction. The level repulsion between these bulk bands and surface states causes surface bands bending down in \Gbar\Mbar\ direction, leading to the observed hexagonal Fermi surface. The surface-only $k\cdot p$ Hamiltonian cannot capture this interaction with bulk bands, in contrast to first-principle calculations.\cite{Basak:2011p13107}

The anisotropic behavior of the surface-state dispersions is also reflected in the HWHM of the Lorentzian fits, as shown in Fig. \ref{Anisotropy1}\,(d) for 21  and 55 eV photons, which provide similar results (solid and dotted lines, respectively). In the BE range where the dispersion along the \Gbar\Mbar\ direction starts to deviate from the linear behavior, i.e., from about 210\,meV BE in the present case, the HWHM increases from a value of $\sim$0.011\,\AA$^{-1}$ to a maximum value of $\sim$0.015\,\AA$^{-1}$ at a BE of $\sim$80\,meV and then decreases towards the Fermi energy to approach the value of $\sim$0.009\,\AA$^{-1}$, which approximately matches the one along the \Gbar\Kbar\ direction. If we would assume a constant and isotropic bare group velocity $v_{0}$ for both directions, this would mean enhanced scattering for electrons moving along \Gbar\Mbar\ as compared to electrons moving along \Gbar\Kbar. However, the experimental group velocities are anisotropic and not constant as shown Fig.\,\ref{Anisotropy1}\,(e). We do note that using the definition of the group velocity as ${\partial{E}}/\hbar{\partial{\bf k}}$, implies that it has to be determined from the $E({\bf k})$ dispersion, and not by differentiating the ${\bf k}(E)$ dispersion and then calculating the reciprocal value of the result. This method would be only valid for a linear dispersion where the group velocity is nearly BE-independent. In the case of \BiTe, one has either to approximate the group velocities linearly over small energy intervals, taking into account artificial energy steps, or to determine the $E({\bf k})$ dispersion. For the results shown in Fig. \ref{Anisotropy1}\,(e), we have chosen the latter method and mapped the function $v_g({\bf k})$ into $v_g(E)$. To avoid the introduction of additional noise, we have fitted the resulting functions with polynomials. As seen in Fig. \ref{Anisotropy1}\,(e), along the \Gbar\Kbar\ direction the group velocity varies only slowly with BE and, in particular, it increases towards lower BE following a linear dependence, in agreement with the slight parabolic shape of the TSS dispersion between the Dirac point and the Fermi level. In contrast, along the \Gbar\Mbar\ direction, we find a more pronounced BE dependence of the group velocity which varies between $\sim$3.1\,eV\AA\ and the minimum value of $\sim$1.5\,eV\AA\ reached around a BE of $\sim$80 meV. Specifically, the Fermi velocity $v_F$ is different along the two high-symmetry directions, being $\sim$2.6\,eV\AA\ and $\sim$3.5\,eV\AA\ along \Gbar\Mbar\ and \Gbar\Kbar, respectively.
\begin{figure*}
\centering
\includegraphics [width=0.9\textwidth]{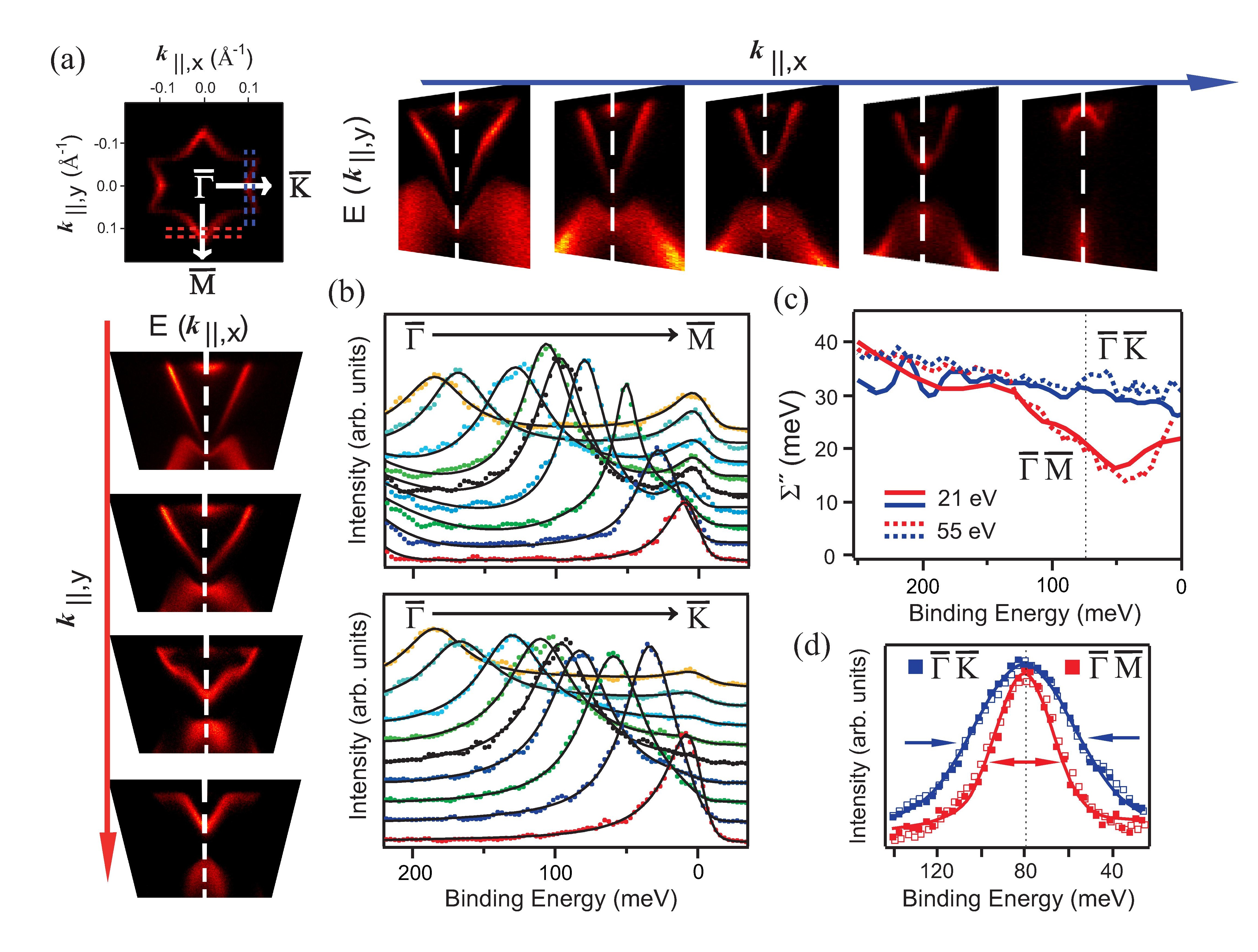}
\caption{(Color online). Extraction of undistorted bands from the constant-energy surface mapping at 21 eV. (a) From cuts perpendicular to the high-symmetry directions a sequence of $E({\bf k}_{\parallel,x})$ ($E({\bf k}_{\parallel,y})$) dispersions at different constant ${\bf k}_{\parallel,y}$ (${\bf k}_{\parallel,x}$) values is obtained (see text). (b) EDCs extracted from the sequence at ${\bf k}_{\parallel,x}$=0 (${\bf k}_{\parallel,y}$=0) along the \Gbar\Mbar\ (\Gbar\Kbar) direction. The EDCs cut through a flat band and show rather symmetric peaks for the surface state, as represented by colored squares. Solid black (dark) lines show the result of the fits. (c) The imaginary part of the self-energy $\Sigma''$ shows the same qualitative behavior as the results obtained from MDCs in Fig.\,\ref{Anisotropy1}\,(f). Solid and dashed lines are results for 21 and 55 eV photons. (d) The differences in the lifetime broadening of the TSS can also be noticed by directly comparing the width of the TSS peaks extracted at a BE of $\sim$80 meV [marked by a vertical dashed-gray (light) line in Fig.\,\ref{Anisotropy2}\,(c)]. Results obtained with $p+s$ linearly (filled symbols) and positive circularly polarized light (open symbols) are shown. In (c) and (d), blue (dark) and red (light) colors denote \GbarKbar\ and \GbarMbar\ directions, respectively.}
 \label{Anisotropy2}
\end{figure*}

If we now use the standard method of multiplying the HWHM values extracted from the fits to MDCs with the experimental group velocities to determine the imaginary part of the self-energy $\Sigma''$, we obtain the results shown in Fig.\,\ref{Anisotropy1}\,(f). We do note that these results are qualitatively reversed as compared to the ones shown in Fig.\,\ref{Anisotropy1}\,(d). After multiplication by the group velocities, we find scattering rates which are almost constant in BE along \Gbar\Kbar, with a slight and smooth decrease while approaching the Fermi level. At a BE higher than $\sim$200 meV, the scattering rates are qualitatively similar for both \Gbar\Kbar\ and \Gbar\Mbar\ directions, as expected for a circular Dirac cone. However, the scattering rates for electrons moving along \Gbar\Mbar\ are strongly reduced as the dispersion experiences more and more the hexagonal deformation. From a BE of $\sim$200 meV towards the Fermi level, an almost linear decrease of the scattering rates is clearly observed along this direction. Below a BE of $\sim$100 meV, the linear decrease slowly saturates, and nearly constant values of $\Sigma''$ are reached down to a BE of $\sim$25\,meV. In the narrow BE window between $\sim$25\,meV and the Fermi level, $\Sigma''$ slightly increases to approach the scattering rates along \Gbar\Kbar, exhibiting a significant upturn behavior in the immediate vicinity of the Fermi level. We point out that in this BE region, a pronounced overlap between the TSS and BCB is clearly observed along \Gbar\Mbar\ with 21 eV photons, as discussed above [see Figs. 2(a) and 2(c)]. One would expect that at 21 eV photon energy, the intensity contribution from the BCB in the overlapping region prevents us from fitting more reliably the TSS peaks in this small BE energy window. However, for 55 eV photons, the BCB is strongly suppressed, and the upturn behavior of $\Sigma''$ along \Gbar\Mbar\ can be clearly observed in Fig.\,\ref{Anisotropy1}\,(f). This finding allows us to attribute the upturn behavior very near the Fermi level to an increased surface-bulk coupling within the overlapping region, which leads to an enhancement of the scattering rate of the surface-state electrons along \Gbar\Mbar. As a result, the scattering rates along \Gbar\Kbar\ and \Gbar\Mbar\ directions behave more isotropically at the Fermi level. It should be emphasized that this type of coupling is independent of the photon energy, despite the fact that the spectral weight of the BCB is strongly suppressed for 55 eV photons. In this respect, we would like to point out the basic fact that the surface-projected bulk band structure, which hosts the signatures of the coupling, is independent of the probing photon energy (effectively \kperp values). In other words, if the overlap between TSS and BCB along \Gbar\Mbar\ direction is observed at a particular photon energy (21 eV in our case), the signatures of enhanced surface-bulk coupling in the linewidth of the TSS will remain photon-energy independent. This will be the case even if the BCB disperses with \kperp or its intensity is suppressed at a particular excitation energy (such as 55 eV in our case). Similarly important, it should be noted that around 80 meV BE, below the BCB minimum, the anisotropy between the scattering rates along \Gbar\Kbar\ and \Gbar\Mbar\ is rather large for both 21 and 55 eV photons. Around this BE, the corresponding constant-energy contours already exhibit a snowflake-like shape with a level of distortion similar to the one of the Fermi surface [see Figs. 2(c) and 2(f)]. This result strongly indicates that away from the immediate vicinity of the Fermi level,  the anisotropic behavior of $\Sigma''$ observed in Fig.\,\ref{Anisotropy1}\,(f) is related to warping. 

We emphasize that the results shown in Fig.\,\ref{Anisotropy1}\,(f) have not been corrected for impurity-scattering contributions, which in a trivial way increase $\Sigma''$ by a small constant and energy-independent value represented by the offset at the Fermi level. Other extra contributions, such as the one from electron-phonon coupling, can be found in the Debye model for the high temperature limit.\cite{Aschcroft} Here the electron-phonon broadening depends linearly on the temperature $T$ as $\Sigma''_{e-ph}$=$\pi\lambda k_{B}T$, where $k_{B}$ and $\lambda$ are the Boltzmann and electron-phonon coupling constants. From temperature-dependent measurements in our own Bi$_2$Te$_3$ samples we find a weak electron-phonon coupling characterized by $\lambda\approx$ 0.17, in agreement with recent studies.\cite{Chen-SciRep-2013} This result leads to $\Sigma''_{e-ph}\approx$ 2 meV, which represents an almost negligible energy-independent contribution to the lifetime broadening due to electron-phonon interaction. Other contributions to the linewidths due to final-state broadening, which for two-dimensional electronic states are reduced to an extra broadening of purely geometrical origin,\cite{Smith-PRB-93,Damascelli-PS-2004} are also negligible. In this case the measured peak widths are compressed or expanded depending on a geometrical factor $C$ so that the measured lifetime broadening is represented by $C\Sigma''$, being $C=1/|1-mv_{0}sin^{2}\theta|$, $v_{0}$ the group velocity, $m$ the electron mass and $\theta$ the corresponding emission angle.\cite{Smith-PRB-93, Damascelli-PS-2004} Due to the small size of the Fermi surface of the TSS, our detail analysis reveals that for 21 and 55 eV photons $C\approx$ 1 along both \Gbar\Mbar\ and \Gbar\Kbar\ directions. Therefore, our results of the MDCs analysis show the importance of a proper group velocity determination and mainly depend on its specific behavior, in particular taking into account the obvious differences between the HWHM and the $\Sigma''$ values obtained in Figs.\,\ref{Anisotropy1}\,(d) and \,\ref{Anisotropy1}\,(f). Hence, in order to confirm the validity of these results and, before discussing the behavior of $\Sigma''$ in more detail, in the following we want to justify the usage of the experimental group velocities by a different method, the results of which are shown in Fig.\,\ref{Anisotropy2}. 

Differently from the method used to obtain the results of Fig.\,\ref{Anisotropy1}, in Fig.\,\ref{Anisotropy2} we perform an analysis of the TSS spectral width using energy distribution curves (EDCs) extracted from the ARPES dispersions shown in Figs. 2(a) and 2(d), where the \Gbar\Kbar\ and \Gbar\Mbar\ directions are oriented along ${\bf k}_{\parallel,x}$ and ${\bf k}_{\parallel,y}$ wave vectors, respectively (i.e., parallel and perpenducular to the analyzer entrance slit). This method allows us, on the one hand, to avoid an artificial broadening caused by the angle between the band dispersion and the cut direction and, on the other hand, to check whether the sample geometry influences the resulting scattering rates by directly comparing to the results of Fig.\,\ref{Anisotropy1}. Moreover, if the self-energy $\Sigma({\bf k},E)$ is slowly varying with BE, the HWHM of the TSS peaks in each EDC gives the imaginary part of the self-energy $\Sigma''$. This method has been used, for instance, to study many-body effects in Mo(110) surfaces,\cite{Valla:1999p13376} 3d ferromagnets,\cite{Sanchez-Barriga-PRL-2009,Sanchez-Barriga-PRB-2010,Sanchez-Barriga-PRB-2012} or high-$T_{c}$ superconductors.\cite{Eisaki-PRL-2008} 

The procedure we have used is illustrated in detail in Fig.\,\ref{Anisotropy2}. At the top-left side of Fig.\,\ref{Anisotropy2}\,(a), as a reference, we show a constant-energy surface extracted at a BE of $\sim $80\,meV from the 21 eV data of Fig. 2. In order to quantitatively analyze the peak widths corresponding to the TSS bands along both \Gbar\Mbar\ and \Gbar\Kbar\ directions using EDCs, we extract two sequences of $E({\bf k}_{\parallel})$ plots from the 21 eV volumetric data set shown in Fig. 2(a). These sequences cut perpendicularly to the \Gbar\Mbar\ and \Gbar\Kbar\ directions, as indicated on top of the constant-energy surface of Fig.\,\ref{Anisotropy2}\,(a) by dashed-red (light) and blue (dark) lines oriented parallel to ${\bf k}_{\parallel,x}$ and ${\bf k}_{\parallel,y}$ wave vectors, respectively. Few-selected results of the $E({\bf k}_{\parallel})$ plots belonging to each sequence are shown on the right and bottom sides of Fig.\,\ref{Anisotropy2}\,(a). Note that along \Gbar\Mbar, the plots are displayed from the top to the bottom of Fig.\,\ref{Anisotropy2}\,(a), while along \Gbar\Kbar, from left to right, which corresponds to increasing values of ${\bf k}_{\parallel,y}$ and ${\bf k}_{\parallel,x}$, respectively. In each of these plots, the TSS band reaches a minimum which is represented by a small energy region located at ${\bf k}_{\parallel,x}$=0 (${\bf k}_{\parallel,y}$=0) along \Gbar\Mbar\ (\Gbar\Kbar). Clearly, these band minima change in BE position through the sequences. In order to obtain the imaginary part of the self-energy $\Sigma''$, we analyze the peak widths of these band minima along \Gbar\Kbar\ (\Gbar\Mbar) using the corresponding EDCs extracted at ${\bf k}_{\parallel,y}$=0 (${\bf k}_{\parallel,x}$=0) [marked by dashed-white (light) lines in each $E({\bf k}_{\parallel})$ plot]. This way, we ensure that the broadening of the peaks does not suffer from the aforementioned artificial broadening. Moreover, because the geometrical factor $C$ is $\approx$ 1 along both \Gbar\Mbar\ and \Gbar\Kbar\ directions, we do not expect extra broadening of purely geometrical origin contributing to the EDC widths. In Fig.\,\ref{Anisotropy2}\,(b), we show few-selected fits to EDCs extracted for 21 eV photons from the sequence at ${\bf k}_{\parallel,x}$=0 (${\bf k}_{\parallel,y}$=0) along the \Gbar\Mbar\ (\Gbar\Kbar) direction [top (bottom) panel]. The experimental EDCs, which are represented by colored-dotted lines, exhibit rather symmetric peaks following the energy-momentum dispersion of the TSS. Each color corresponds to a different cut through the sequences. In analogy to the method used in Fig.\,\ref{Anisotropy1}, the peaks are fitted by Lorentzian profiles convoluted with a Gaussian function representing the energy resolution (see, for instance, Ref. \onlinecite{Sanchez-Barriga-PRB-2010} for more details on the procedure). Similar analysis was performed for the 55 eV data of Fig. 2(d). Although we find almost absent asymmetries caused by the BE dependence of $\Sigma''$, other problems connected to the analysis of the EDCs might still exist, like the secondary electron background or deviations near the cut-off at the Fermi level. Nevertheless, in our case we find a low-background signal and rather symmetric peaks across the Fermi level, in a similar fashion as in previous studies.\,\cite{Valla:1999p13376, Valla:2012p13194,Pan-PRL-2012, Chen-SciRep-2013}

The $\Sigma''$ values, which are directly represented by the HWHM of the Lorentzian peaks obtained by fitting the corresponding EDCs, are shown in Fig.\,\ref{Anisotropy2}\,(c) for both \Gbar\Kbar\ [blue (dark) circles] and \Gbar\Mbar\ [red (light) circles] high-symmetry directions. The analysis for 21 and 55 eV photons provides similar results (solid and dotted lines, respectively). Again, pronounced differences in the behavior of $\Sigma''$ along \Gbar\Mbar\ and \Gbar\Kbar\, leading to a remarkable momentum dependence of the scattering rates, are clearly observed. Such differences in the lifetime broadening of the TSS are independent of the light polarization, as it can also be noticed in Fig.\,\ref{Anisotropy2}\,(d) where we directly compare for linearly and circularly polarized 21 eV photons of positice helicity the width of the TSS peaks extracted at a BE of $\sim$80 meV [marked by a vertical dashed-gray (light) line in Fig.\,\ref{Anisotropy2}\,(c)]. Note that this BE corresponds to the constant-energy contour shown in the top-left side of Fig.\,\ref{Anisotropy2}(a), where the hexagonal distortion of the constant-energy surfaces and the anisotropy of the scattering rates are rather large. We point out that similar results are obtained either by reversing the light helicity of the circular polarization to negative or using s-polarized light. Our results are consistent with unobservable differences between the orbital-projected scattering rates and with the fact that the TSS signal within the first few atomic layers is dominated by a strong contribution from $p_{z}$-type orbitals. Moreover, comparing Figs.\,\ref{Anisotropy1}(f) and \,\ref{Anisotropy2}\,(c), we find qualitative agreement, a fact that corroborates our initial observation of reduced scattering rates along \Gbar\Mbar\ as compared to \Gbar\Kbar\ direction in the presence of hexagonal warping. Hence, it is reasonable to assume that for the TSS of \BiTe, multiplication of the HWHM values obtained from MDCs with the experimental group velocities is an equally valid method as the EDC analysis, at least to qualitatively look at the behavior of $\Sigma''$ over a certain BE range. In this respect, we have to stress that the validity of both methods hinges on the assumption that the TSSs are weakly-interacting states. This fact is also evidenced in the EDC analysis by the excellent accuracy of the EDC fits using symmetric functions or, in other words, by the absence of asymmetries in the TSS peaks due to damping of quasiparticle excitations. Additionally, the qualitative agreement between Figs.\,\ref{Anisotropy1}(f) and \,\ref{Anisotropy2}\,(c) implies that our results are independent of the sample geometry, indicating that other effects such as orbital spin interference\cite{Zhu-PRL-2013} do not play a role in our observation of anisotropic scattering rates. Despite the fact that very near the Fermi level the EDC analysis becomes more difficult due to the peak intensity cut-off by the Fermi energy, it is also important to observe that along \Gbar\Mbar\ direction, the results of Fig.\,\ref{Anisotropy2}\,(c) reveal a significant upturn behavior of $\Sigma''$ in the immediate vicinity of the Fermi level. This observation is in qualitative agreement to the results of the MDC analysis shown in Fig.\,\ref{Anisotropy1}\,(f), indicating that the upturn behavior and its independence on the photon energy are not connected to the sample geometry or to the analysis method, but rather to surface-bulk coupling, as discussed above. 

In the following, we would like to focus on the main outcome of the present work which is the anisotropic effect of warping on the scattering rates in the region where the TSS and the BCB are decoupled. In particular, we would like to examine various mechanisms that can explain this observation. These mechanisms, which are all related to warping and in some degree connected to each other, contribute with different weights, as we will discuss below.
\begin{figure*}
\begin{center}
\includegraphics[width=0.9\textwidth]{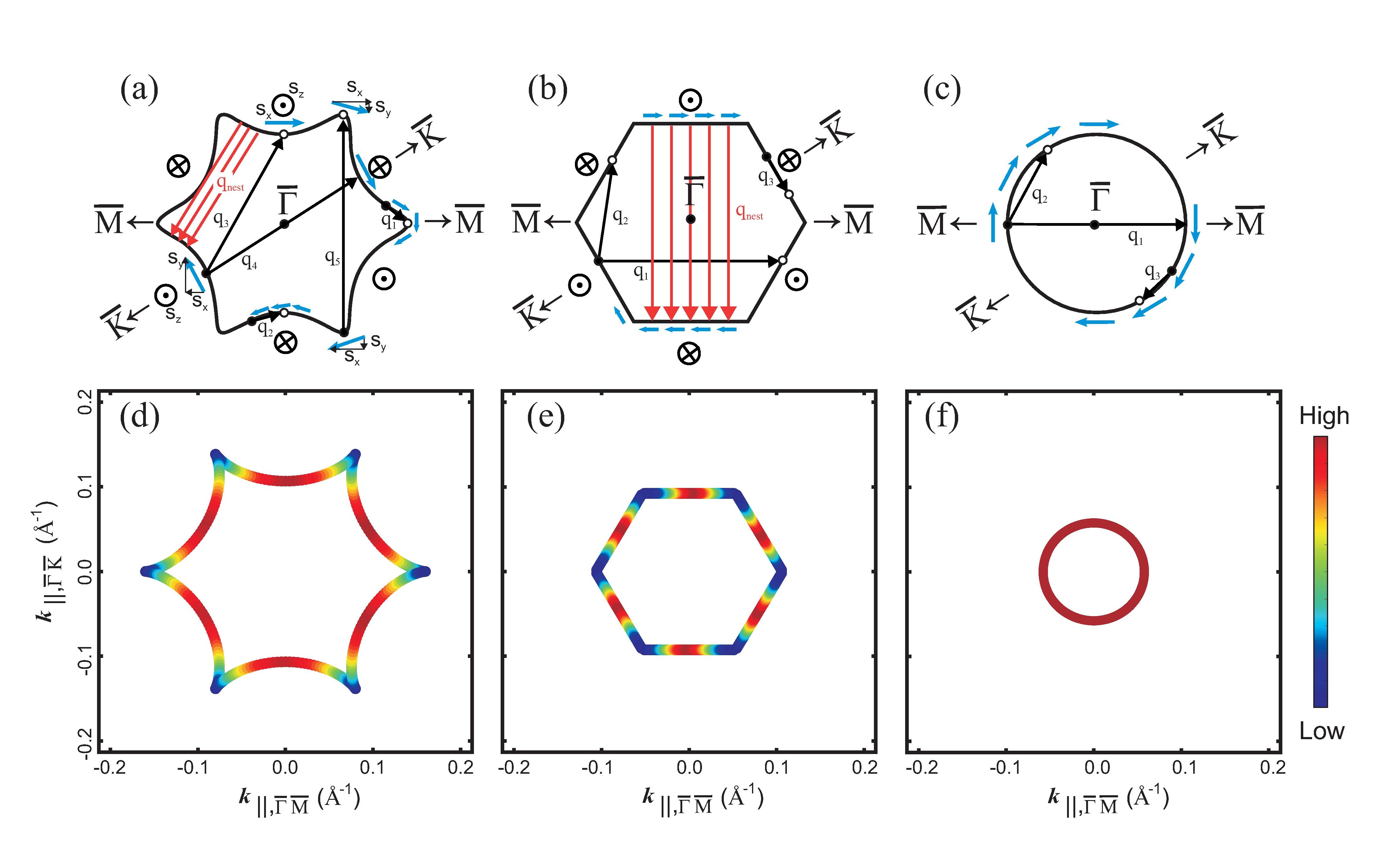}
\end{center}
\caption{(Color online). (a)-(c) Schematics of different scattering channels. (a) For a snowflake-like constant-energy contour, the in-plane spin directions [components $s_x$ and $s_y$, as indicated by blue (light) arrows] change stronger in the vicinity of the \Gbar\Mbar\ direction as compared to the \Gbar\Kbar\ direction, leading to a reduced probability of small-angle scattering events, e.g., $q_1$ as compared to events exemplified by $q_2$. The out-of-plane spin rotation for ${\bf k}$ vectors away from the \Gbar\Mbar\ direction is sixfold symmetric, as indicated by the symbols $\odot$ ($s_z$ points out of the paper plane) and $\otimes$ ($s_z$ points into the paper plane). The scattering probability is increased for scattering under $q_3$, whereas backscattering (e.g., $q_4$) is still forbidden. The almost flat regions of the constant-energy contours, connected by red (light) scattering vectors ($q_{nest}$) fulfil a near nesting condition. (b) Even stronger nesting 
is expected for a hexagonal shape, however, for this case time-reversal symmetry strongly suppresses scattering. Thus a spin-density wave would be favored 
over a charge-density wave (see text). (c) Similar scheme for a circular shape where scattering vectors such as $q_2$ and $q_3$ contribute isotropically to the total scattering rate. (d)-(f) Simplified model calculations of the spin-dependent scattering amplitude, which includes contributions from all possible {\bf q} vectors within (d) a snowflake-like, (e) a hexagon-shaped and (f) a circular constant-energy contour.}
\label{Anisotropy3a}
\end{figure*}
An important mechanism contributing to the anisotropic behavior of the scattering rates is spin-dependent scattering, i.e., spin-dependent decay of photoholes in our case. For a two-dimensional, non-degenerated, time-reversal-symmetric surface state, backscattering is forbidden, and the presence of strong warping does not affect this prohibition. However, this scenario is only valid for 180$^\circ$ backscattering, as scattering under different angles is not completely suppressed. In particular, the probability for a scattering event away from 180$^\circ$ will be enhanced if there is a strong overlap between parallel spin configurations in the initial and final states involved in every scattering event.\cite{Roushan} Equally important, as we have already mentioned, is the theoretical prediction that warping affects the spin such that it is not anymore locked perpendicular to the electron momentum, but tangential to the constant-energy contours.\,\cite{Basak:2011p13107} Such predicted ground-state electron spin texture is illustrated in Fig.\,\ref{Anisotropy3a}, for a snowflake-like [Fig.\,\ref{Anisotropy3a}\,(a)], a hexagon-shaped  [Fig.\,\ref{Anisotropy3a}\,(b)], and a circular [Fig.\,\ref{Anisotropy3a}\,(c)] constant-energy contour. Note that due to the limited momentum resolution of presently available SR-ARPES setups, a current experimental challenge is the accessibility to the pronounced tangential in-plane spin rotations seen in Fig.\,\ref{Anisotropy3a}\,(a) near the \Gbar\Mbar\ direction. This type of tangential spin rotations have been claimed to be found experimentally in \BiSe,\cite{Wang:2011p13090} where the Dirac cone is only slightly distorted by warping, but the method used in these experiments, i.e., circular dichroism, appears questionable as it reverses several times with photon energy.\cite{Sanchez-Barriga-PRX-2014, Scholz-PRL-2013} 

In the context of SR-ARPES, it should be emphasized that the recently observed changes in the spin polarization of photoelectrons emitted from TSSs as a function of light polarization are due to the interplay between spin-dependent matrix-elements, dipole-selection rules and photoelectron spin-interference effects. \cite{JozwiakNP13, Sanchez-Barriga-PRX-2014, Zhu-PRL-2013} In other words, the observed effect of photoelectron spin manipulation with light polarization is not caused by light-induced changes in the TSS initial-state spin texture. \cite{JozwiakNP13, Sanchez-Barriga-PRX-2014, Zhu-PRL-2013} Moreover, our observation in Fig.\,\ref{Anisotropy2}\,(d) that the anisotropy of the scattering rates does not change with the photon polarization demonstrates that it is completely independent of light-induced changes in the photoelectron spin. Therefore, in the following we would like to focus our attention in the theoretically predicted ground-state spin texture of a strongly warped Dirac cone. \cite{Basak:2011p13107,Yazyev-PRL-2010} This peculiar spin texture allows us to explain in detail our observation of anisotropic scattering rates in \BiTe\ via spin-dependent scattering, as we will discuss below. 
   
Let us assume that an electron has been removed from the vicinity of a tip of the snowflake-shaped constant-energy surface [Fig.\,\ref{Anisotropy3a}\,(a)] along the \Gbar\Mbar\ direction. In this case, the electron spin will be fully parallel to the surface plane (i.e., $s_x\ne0$ and/or $s_y\ne0$, while $s_z=0$), consistent with the mirror symmetry of the crystal.\,\cite{Basak:2011p13107} Scattering under small angles is allowed, but it is less likely to occur along the \Gbar\Mbar\ direction [e.g., $q_1$ in Fig.\,\ref{Anisotropy3a}\,(a)], due to the fact that the spin varies strongly while following the constant-energy con tour [as indicated by blue (dark) arrows in Fig.\,\ref{Anisotropy3a}]. In contrast, a hole caused by removing an electron from the same BE with its linear momentum pointing along the \Gbar\Kbar\ direction can be filled with electrons from adjacent sites in ${\bf k}$-space, since the in-plane spin components are nearly parallel to each other [e.g., $q_2$ in Fig.\,\ref{Anisotropy3a}\,(a)]. Furthermore, along the \Gbar\Kbar\ direction, the electron spin exhibits a finite out-of-plane component $s_z$ [denoted by $\odot$ and $\otimes$ symbols in Figs. \,\ref{Anisotropy3a}(a) and \,\ref{Anisotropy3a}(b)]. This spin component, which can lead to out-of-plane spin polarization values expected to be as large as $\sim$30\%,\,\cite{Basak:2011p13107,Yazyev-PRL-2010} reverses its sign with an angular periodicity of $\frac{\pi}{3}$ and vanishes along the \Gbar\Mbar\ direction due to mirror symmetry.\,\cite{Basak:2011p13107} Note that $s_z$ varies only smoothly around the distorted constant-energy contours of Figs.\,\ref{Anisotropy3a}(a) and \,\ref{Anisotropy3a}(b), meaning that even though adjacent sites in ${\bf k}$-space might have different absolute values of the out-of-plane spin polarization, these values will be always of the same sign. This specific behavior of $s_z$ strongly favours spin-dependent scattering as explanation for the increased scattering rates along \Gbar\Kbar\ direction. Such an argument can be further exemplified by considering a scattering vector like $q_3$ in Fig.\,\ref{Anisotropy3a}\,(a). The initial state (marked by the black (dark) dot at the beginning of $q_3$) has finite spin components $s_x$ and $s_y$, but there is no overlap between these spin components and the ones in the final state, which has $s_y$=0 and $s_x$ pointing antiparallel with respect to $s_x$ in the initial state. However, $s_z$ is pointing out of the surface plane for both initial and final states, leading to a relatively strong contribution to the scattering probability due to the large overlap between these two parallel out-of-plane spin configurations.\cite{Roushan} In contrast, a comparable scattering event for electrons moving along the \Gbar\Mbar\ direction, such as the one exemplified in Fig.\,\ref{Anisotropy3a}\,(a) by $q_5$, where $s_z$=0, will have only a small overlap between the corresponding $s_y$ in-plane spin components in both initial and final states. Similarly, in a hexagon-shaped constant-energy contour there will be always a small overlap between parallel spin components for scattering vectors such as $q_1$, $q_2$ or $q_3$ [see Fig. \,\ref{Anisotropy3a}(b)]. In the case of a circular constant-energy contour, because $s_z$=0, only contributions from scattering vectors such as $q_2$ or $q_3$ would be possible [see Fig.\,\ref{Anisotropy3a}\,(c)].

In order to gain more insight on the overlap between different spin configurations, we have evaluated on the basis of a simplified model spin-dependent scattering matrix elements of the form $|\langle {\bf s}({\bf k})\,|\,{\bf s}({\bf k}+{\bf q})\rangle|^2$. These matrix elements represent the scattering amplitude between two different spin states {\bf s}({\bf k}) and {\bf s}({\bf k}+{\bf q}). The contribution to the total scattering is represented in our calculation by the sum of individual matrix elements for all possible {\bf q} vectors around the constant-energy contour, weighted to the total spin polarization. Here ${\bf s}$ denotes the electron initial-state spin configuration in three dimensions, which is represented by the predicted ground-state spin texture of a strongly warped Dirac cone.\cite{Basak:2011p13107, Yazyev-PRL-2010} This type of spin-selective scattering processes are enhanced for aligned spins and completely suppressed for opposite and perpendicular spin orientations.\cite{Roushan} In our simplified model the constant-energy contours are modeled by parametric equations, and the scattering amplitude in each {\bf k}-space site is quantified by the inner product of three-dimensional spin vectors. Scattering events with zero or anti-parallel spin projections are strictly forbidden. The spin texture is represented by spin-polarization vectors with in-plane spin components tangential to the constant-energy contours. The magnitude of the out-of plane spin polarization along the \Gbar\Kbar\ direction is fixed to the theoretically expected value of $\sim$30\%,\cite{Basak:2011p13107, Yazyev-PRL-2010} and interpolated all around the constant-energy contour using a sinusoidal function with a periodicity of 2$\pi$/3. Similarly, we consider a constant magnitude of 50\% for the expected value of the total initial-state spin polarization of the TSS. \cite{Basak:2011p13107, Yazyev-PRL-2010}  Other effects, such as orbital spin interference,\cite{Zhu-PRL-2013} are not taken into account. 

The results of the model calculations are depicted in Figs.\,\ref{Anisotropy3a}\,(d), \,\ref{Anisotropy3a}\,(e), and \,\ref{Anisotropy3a}\,(f) for a snowflake-like, a hexagon-shaped and a circular constant-energy contour, respectively. Our simulation reveals that in the case of an ideal circle, spin-dependent scattering events are likely to occur isotropically. When the influence of warping deforms the constant-energy contour from a hexagon to a snowflake-like shape, the anisotropy in the scattering is manifested as higher and lower spin-dependent scattering amplitudes along the \Gbar\Kbar\ and \Gbar\Mbar\ directions, respectively. These results are in qualitative agreement with our experimental finding of anisotropic scattering rates, indicating that spin-selective scattering is the driving force behind this observation. We point out that including in our model unpolarized BCB states which only overlap with the TSS bands along the \Gbar\Mbar\ direction, leads to a nearly isotropic scattering behavior. This is the case if we assume that enhanced surface-bulk coupling eventually quenches the spin polarization in the overlapping region. Such a strong reduction of the spin polarization along \Gbar\Mbar\ opens up multiple scattering channels along all directions. This limiting case of our simplified model is in principle consistent with the upturn behavior of $\Sigma''$ along the \Gbar\Mbar\ direction.

Another mechanism that might play an important role is nesting.\,\cite{Fu:2009p10385} In contrast to an ideally circular Dirac cone, or for example a two-dimensional free-electron gas, the warped constant-energy contours of \BiTe\ in the vicinity of the \Gbar\Kbar\ direction fulfil a nesting condition, especially those of hexagonal shape, such as the one shown in Fig.\,\ref{Anisotropy3a}\,(b). Note that for a snowflake-like constant-energy contour, at least a near-nesting condition must also be fulfilled.\,\cite{Hasan:2009p10386} Nesting means that parts of the constant-energy contour can be mapped onto each other by a single scattering vector $q_{nest}$ [indicated by the red (light) arrows connecting opposite sides of the constant-energy contours in Fig.\,\ref{Anisotropy3a}\,(a) and \,\ref{Anisotropy3a}\,(b)]. It is expected that fulfilling the nesting leads to enhanced scattering, but also to instabilities, like, for example, a charge-density wave. However, time-reversal symmetry disfavours the formation of a charge-density wave, \cite{Fu:2009p10385} because independently of the warping, states at opposite ${\bf k}_{\parallel}$ momenta have opposite spin [see, e.g., the blue (dark) arrows in Fig.\,\ref{Anisotropy3a}\,(b)]. Alternatively, the formation of a spin-density wave in \BiTe\ has been also suggested as a possible instability introduced by the warping.\,\cite{Fu:2009p10385, Hasan:2009p10386} Although we do not observe clear signs of a spin-density wave in the ARPES band dispersions of Fig. 2, i.e., a band gap at the Fermi level, this fact does not completely exclude a contribution from nesting in the enhancement of the scattering rates along the \Gbar\Kbar\ direction. However, since the nesting vector $q_{nest}$ given in Fig.\,\ref{Anisotropy3a}\,(b) is strongly suppressed by time-reversal symmetry and the one in Fig.\,\ref{Anisotropy3a}\,(a) does not favour scattering of electrons along the \Gbar\Kbar\ direction, it is likely that nesting plays minor role in our observation of anisotropic scattering rates.

Finally, scattering of photoholes through transitions into the bulk continuum might additionally contribute to the scattering rates because of the presence of dispersing bulk states which do not directly overlap with the TSS bands. In an analysis of the scattering rates of the surface-state electrons of \BiSe,\,\cite{Park:2010p10834} a pronounced decrease of $\Sigma''$ was found from the minimum of the BCB towards the Fermi level along the \Gbar\Kbar\ direction. This observation was attributed to scattering processes strongly dominated by transitions into bulk electronic states, a finding that is only in qualitative agreement with the overall behaviour of our data for \BiTe\ along the \Gbar\Kbar\ direction. Note that we observe average values of $\Sigma''$ that change very smoothly and by a small amount between the minimum of the BCB and the Fermi level along this direction. Moreover, the upturn behaviour of $\Sigma''$ in the immediate vicinity of the Fermi level along the \Gbar\Mbar\ direction shows the opposite trend that the one expected from this type of scattering channels. Likewise, in a BE range around the Dirac point, parts of the TSS are surrounded by the BVB and photoholes can also be scattered through transitions into the bulk continuum. As a result, a strong decrease of the scattering rates would be equally expected from high BE up to around the BVB maximum. However, our results reveal that in this BE region the changes in $\Sigma''$ are not significant. All these observations indicate that quasiparticle decay through transitions into the bulk continuum is likely to play a minor role in the anisotropy of the scattering rates.

It is often stated that an ideal Dirac cone prevents electrons from backscattering. However, as the constant-energy contours inside the bulk band gap become more and more circular with increasing BE, in Figs. \,\ref{Anisotropy1}(f) and \,\ref{Anisotropy2}(c) we observe a pronounced increase of the scattering rates along \Gbar\Mbar\ as compared to \Gbar\Kbar\ direction. This effect is due to the fact that increasing the BE inside the bulk band gap leads to a higher probability for small-angle scattering events because the variation of the spin direction in adjacent ${\bf k}$-points is smaller in a circle or a hexagon than at the tips of a snowflake-like constant-energy contour, as discussed above. The overall effect is consistent with our explanation that spin-dependent scattering causes the observed anisotropy of the scattering rates in \BiTe.

To summarize, we have analyzed the effect of strong hexagonal warping on the lifetime broadening of the TSS of the prototype TI \BiTe\ by means of angle-resolved photoemission. We have shown that the large Fermi-surface distortion introduced by warping leads to an anisotropy of the scattering rates which is larger along the \Gbar\Kbar\ high-symmetry direction. Based on the theoretically predicted behavior of the ground-state spin texture of a strongly warped Dirac cone, we have identified spin-dependent scattering as the underlying mechanism giving rise to anisotropic scattering rates of the surface-state electrons. These results could help paving the way for controlling  surface-scattering processes via external gate voltages in future spintronic devices based on TIs.

\end{document}